\documentclass[useAMS,usenatbib]{mnras}

\input{epsf}
\usepackage{natbib}
\usepackage{graphicx}	
\usepackage{deluxetable}
\usepackage{amsmath}
\usepackage{longtable}
\usepackage{lscape}
\usepackage{threeparttable}
\usepackage{threeparttablex}
\usepackage[singlelinecheck=false]{caption}

\newcounter{species}
\def\ion#1#2{\setcounter{species}{#2}#1$\;${\sc\roman{species}}\relax}

\newcommand{\Msun}{M$_{\odot}$}

\newcommand{\OIIIdblt}{[O\,{\sc iii}]~$\lambda\lambda$4959, 5007}
\newcommand{\OIIdblt}{[O\,{\sc ii}]~$\lambda\lambda$3726, 3729}

\newcommand{\SIIdblt}{[S\,{\sc ii}]~$\lambda\lambda$6716, 6731}
\newcommand{\NIIdblt}{[N\,{\sc ii}]~$\lambda\lambda$6548, 6583}
\newcommand{\NeIIIdblt}{[Ne\,{\sc iii}]~$\lambda\lambda$3869, 3967}

\newcommand{\OIw}{[O\,{\sc i}]~$\lambda$6300}

\newcommand{\Hb}{{H}$\beta$}
\newcommand{\Ha}{{H}$\alpha$}
\newcommand{\FeII}{Fe\,{\sc ii}}

\newcommand{\NII}{[N\,{\sc ii}]}

\newcommand{\OI}{[O\,{\sc i}]}
\newcommand{\OII}{[O\,{\sc ii}]}
\newcommand{\OIII}{[O\,{\sc iii}]}

\newcommand{\SII}{[S\,{\sc ii}]}

\newcommand{\NeIII}{[Ne\,{\sc iii}]}

\def\lsim{\lower0.3em\hbox{$\,\buildrel <\over\sim\,$}}
\def\gsim{\lower0.3em\hbox{$\,\buildrel >\over\sim\,$}}

\def\sun{\hbox{$\odot$}}

\def\lesssim{\mathrel{\hbox{\rlap{\hbox{%
 \lower4pt\hbox{$\sim$}}}\hbox{$<$}}}}
\def\gtrsim{\mathrel{\hbox{\rlap{\hbox{%
 \lower4pt\hbox{$\sim$}}}\hbox{$>$}}}}

\def\farcs{\hbox{$.\!\!^{\prime\prime}$}}

\title[The disappearing central engine of J1011+5442]{Now You See It, Now You Don't: The Disappearing Central Engine of the Quasar J1011+5442}

\author[J. C. Runnoe et al.]
{\parbox{\textwidth}{Jessie C. Runnoe$^{\,1},$\thanks{E-mail: \texttt{runnoejc@psu.edu}}
Sabrina Cales$^{\,2,3}$,
John J. Ruan$^{\,4}$,
Michael Eracleous$^{\,1}$, 
Scott F. Anderson$^{\,4}$,
Yue Shen$^{\,5,6}$\thanks{Hubble Fellow},
Paul Green$^{\,7}$,
Eric Morganson$^{\,7}$, 
Stephanie LaMassa$^{\,2}$,
Jenny E. Greene$^{\,8}$,
Tom Dwelly$^{\,9}$,
Donald P. Schneider$^{\,1}$,
Andrea Merloni$^{\,9}$,
Antonis Georgakakis$^{\,9}$}
\vspace{0.4cm}\\ \\
\parbox{\textwidth}{$^{1}$Department of Astronomy \& Astrophysics, and Institute for Gravitation and the Cosmos, The Pennsylvania State University, 525 Davey Lab, University Park, PA 16802, USA\\
$^{2}$Yale Center for Astronomy \& Astrophysics, Physics Department, P.O. Box 208120, New Haven, CT 06520, USA \\
$^{3}$Department of Astronomy, Faculty of Physical and Mathematical Sciences, Universidad de Concepci\'{o}n, Casilla 160 C, Concepci\'{o}n, Chile \\
$^{4}$Department of Astronomy, University of Washington, Box 351580, Seattle, WA 98195, USA \\
$^{5}$Carnegie Observatories, 813 Santa Barbara Street, Pasadena, CA 91101, USA \\
$^{6}$Department of Astronomy, University of Illinois at Urbana-Champaign, Urbana, IL 61801, USA \\
$^{7}$Harvard Smithsonian Center for Astrophysics, 60 Garden St, Cambridge, MA 02138, USA\\
$^{8}$Princeton University Observatory, Peyton Hall, Princeton, NJ 08544, USA\\
$^{9}$Max-Planck-Institut f\"ur Extraterrestrische Physik (MPE), Giessenbachstrasse 1, D-85748, Garching bei M\"unchen, Germany
}}

\begin{document}		

\date{Preprint 2015 March 7}

\pagerange{\pageref{firstpage}--\pageref{lastpage}} \pubyear{2015}

\maketitle

\label{firstpage}

\begin{abstract}
We report the discovery of a new ``changing-look'' quasar, SDSS J101152.98+544206.4, through repeat spectroscopy from the Time Domain Spectroscopic Survey. This is an addition to a small but growing set of quasars whose blue continua and broad optical emission lines have been observed to decline by a large factor on a time scale of approximately a decade. The 5100~\AA\ monochromatic continuum luminosity of this quasar drops by a factor of $> 9.8$ in a rest-frame time interval of $< 9.7$ years, while the broad H$\alpha$ luminosity drops by a factor of 55 in the same amount of time. The width of the broad H$\alpha$ line increases in the dim state such that the black hole mass derived from the appropriate single-epoch scaling relation agrees between the two epochs within a factor of 3. The fluxes of the narrow emission lines do not appear to change between epochs. The light curve obtained by the Catalina Sky Survey suggests that the transition occurs within a rest-frame time interval of approximately 500 days. We examine three possible mechanisms for this transition suggested in the recent literature. An abrupt change in the reddening towards the central engine is disfavored by the substantial difference between the timescale to obscure the central engine and the observed timescale of the transition. A decaying tidal disruption flare is consistent with the decay rate of the light curve but not with the prolonged bright state preceding the decay; nor can this scenario provide the power required by the luminosities of the emission lines. An abrupt drop in the accretion rate onto the supermassive black hole appears to be the most plausible explanation for the rapid dimming. 
\end{abstract}


\section{Introduction}
The dichotomy between Type 1 and Type 2 active galactic nuclei (AGN) is among the most studied in astrophysics \citep[for a review, see][]{netzer15}.  The classes are empirically determined; Type 1 AGN are characterized by having both broad (on the order of $10^{3}\,$km~s$^{-1}$) as well as narrow (on the order of $10^{2}\,$km~s$^{-1}$) emission lines, whereas only the narrow lines are present in the spectra of Type 2 AGN.  The intermediate Type 1.8 and 1.9 classifications describe objects with weak or absent broad \Hb\ emission, respectively, but a broad \Ha\ line \citep{osterbrock81}.  In the simplest scenario, these objects can all be unified by orientation \citep{antonucci93}; Type 2 AGN are viewed edge-on, with an equatorial dusty structure obscuring the line of sight down to the broad-line region (BLR), while Type 1 objects have an unobscured, face-on view.  Early support for this scheme came from ground-breaking views of broad emission lines in Type 2 objects observed in scattered light \citep{antonucci85}.  In this picture, intermediate types are explained by partial obscuration or reddening by optically thin dust \citep{stern12b}.  In all cases, the narrow emission lines originate farther from the central engine in the narrow-line region (NLR), and are not hidden.  

While the simplicity of the above orientation-based scheme is compelling, many have suggested that it does not fully explain the Type 1-Type 2 dichotomy.  An alternative possibility is that the link between classifications might be evolutionary \citep[e.g,][]{penston84}.  Based on NGC~4151 \citep[see also][]{lyutyj84} and 3C~390.3, which were both initially observed to have broad emission lines that later disappeared, Type 2 AGNs may be Type 1s that have ``turned off''.  Variations along this line invoke age, environment, or luminosity effects and are summarized by \citet{gaskell14} \citep[see also,][]{villarroel14}.		
		
Objects that transition between Type 1 and 2 provide a unique opportunity to study the evolutionary side of unification.  Although not common, the number of known objects in this class has been growing.  The term ``changing-look'' was initially used to describe X-ray AGN with variable obscuration \citep[e.g.,][]{matt03,puccetti07,bianchi09,risaliti09,marchese12}, but has lately widened to include objects that transition between Types 1 and 2 (an optical classification).  There are a handful of objects like NGC 4151 and 3C~390.3, where the broad emission lines and quasar continuum fade or disappear completely below the detection limit \citep{collin73,tohline76,sanmartim14,denney14,barth15s} as well as instances where the transition goes in the other direction and the quasar ``turns on'' \citep{cohen86,storchi-bergmann93,aretxaga99,eracleous01,shappee14}.  The timescales for the transition are typically found to be on the order of years, and obscuration or changes in accretion rate are often the preferred explanation.  However, these objects have all been at notably low redshift and luminosity.

\citet{lamassa15} report the discovery of the first changing-look quasar, the most luminous and distant observed to date, with a redshift of $z=0.31$.  SDSS J015957.64+003310.5 (hereafter J0159+0033), is an X-ray selected AGN from the Stripe 82X survey \citep{lamassa13a,lamassa13b} that transitioned from a Type 1 to a Type 1.9 within a $\sim9$ year period.  Initially observed in the first data release of the Sloan Digital Sky Survey \citep[SDSS DR1,][]{abazajian03_short}, J0159+0033 appears as a typical, Type 1 broad-line quasar.  In 2010 it was re-observed in the Baryon Oscillation Spectroscopic Survey \citep[BOSS,][]{dawson13_short} of the Sloan Digital Sky Survey - III \citep[SDSS-III,][]{eisenstein11s}, revealing a galaxy spectrum with only a weak quasar continuum and residual broad \Ha.  Archival X-ray observations of its bright and dim states show that the dimming in the X-ray flux is consistent with the optical continuum, with no strong evidence for obscuration in either band.  Although variable obscuration cannot be emphatically ruled out, \citet{lamassa15} favor an interpretation where the changes in J0159+0033 are due to a decrease in accretion power.

The discovery of luminous analogs of the nearby changing-look AGNs is significant because it demonstrates that such abrupt transitions can occur in quasars where the black holes are more massive and the physical time scales correspondingly longer than their low-redshift counterparts. Motivated by the above considerations, Ruan~et~al. (submitted) performed an archival search of the SDSS DR12 spectroscopic database, in an effort to find more such quasars.  Their criteria target objects at all redshifts that exhibit significant changes in repeat SDSS observations, either transitioning from quasar to galaxy or vice versa.  In addition to J0159+0033, which is recovered, they identify two new changing-look quasars.  In both cases, the changes in the spectrum are best explained by changes in the luminosity of a pre-existing accretion flow rather than tidal disruption events (TDEs), primarily because the observed narrow-line ratios are indicative of long-term AGN activity. Thus it appears that these are examples of a previously unappreciated mode of quasar variability: prolonged ``on-'' and ``off-states.'' This type of variability may correspond to the bursting behavior found in simulations, such as those of \citet{novak11}, where the accretion rate onto the black hole is intermittent because of the interaction of the quasar outflow with the gas supplied by stellar processes (see also the related discussion by \citealt{schawinski15} and \citealt{merloni15}).  Alternatively, the variability may be the result of thermal instabilities that drive the accretion disk to transition between different physical states \citep{ichimaru77}. Such thermal instabilities may follow from small changes in the accretion rate, but the resulting large, abrupt change in the luminosity in this case would come from the transition of the disk state, e.g., to an advection-dominated flow \citep{narayan94}, rather than a sudden cessation in accretion and draining of the disk.

In this work, we report the discovery of a new changing-look quasar, the first found via visual inspection of the Time-Domain Spectroscopic Survey \citep[TDSS,][]{morganson15s}.  The source, SDSS J101152.98+544206.4 (hereafter J1011+5442), is a $z=0.246$ quasar initially observed for spectroscopy by the SDSS in early 2003 that appeared in the third, fifth, and seventh SDSS quasar catalogs \citep{schneider05,schneider07,schneider10}.  It was targeted again by TDSS in the Few Epoch Spectroscopy (FES) program to study variability of the broad Balmer emission lines in a sample of normal, bright, low-redshift quasars and a second spectrum was obtained in early 2015.  The main goals of this work are to describe the behavior of J1011+5442 and set the stage for a systematic search of TDSS for a sample of changing look AGN.  After collecting existing data for this object (Section~\ref{sec:data}), we evaluate several physical scenarios that might explain the changes in J1011+5442 (Section~\ref{sec:analysis}).  We determine that the changes in spectral state are driven by changes in the accretion power, ruling out variable obscuration and TDEs.  These results are summarized and discussed in the context of previous work in Section~\ref{sec:discussion}.  Throughout this work, we adopt the same cosmology used by \citet{lamassa15} of $H_0 = 70$ km s$^{-1}$ Mpc$^{-1}$, $\Omega_{\Lambda} = 0.73$, and $\Omega_{m} = 0.27$, which leads to a luminosity distance of $D_L = 1.244\;$Gpc and a distance modulus of $\mu=40.47\;$mag for J1011+5442.

\section{Data and Measurements}
\label{sec:data}
\subsection{New and archival data}
For this work, we collect the existing photometry and spectroscopy of J1011+5442.  The bright-state spectrum was taken on January 13, 2003 through a 3\farcs0 diameter fiber and covers the range $3800-9200$~\AA\ with a resolution of $R\sim2000$ \citep{york00,smee13s} and was calibrated with the v5\_3\_12 version of the spectroscopic pipeline.  Imaging and photometry are also available from the SDSS, taken almost a year earlier on February 13, 2002.  The SDSS image of J1011+5442 shows that the host galaxy is extended with a visually asymmetric, potentially disturbed, morphology.  Because the source appears extended, we adopt the Petrosian magnitudes, which use a large aperture (a 3\farcs6 diameter in the $r$ band, corresponding to 21~kpc) to capture virtually all of the light.  These $ugriz$ magnitudes are $18.32\pm0.01$, $18.20\pm0.01$, $17.90\pm0.08$, $17.64\pm0.09$, and $17.52\pm0.03$.  The corresponding absolute magnitude in the $i$ band is $M_i=-22.82$, thus it was included in the SDSS DR10 quasar catalog \citep{schneider10}. From the $ugriz$ colors and the transformations of \citet{jester05} appropriate for quasars of $z<2.1$, we obtain $V=18.01\pm0.01$ and $R=17.64\pm0.01$.  The object was re-observed during the TDSS, a subprogram of the Extended Baryon Oscillation Spectroscopic Survey (eBOSS), on February 20, 2015.  Specifically, J1011+5442 was targeted by an FES program to study the variability of the Balmer lines in high signal-to-noise ratio quasar spectra.  The TDSS spectroscopic fiber had a diameter 2\farcs0 and was centered on a position within 0\farcs1 of the SDSS fiber position.  The new TDSS spectrum covers the range $3600-10000$~\AA\ at $R\sim2000$ and was calibrated with the v5\_7\_8 version of the spectroscopic pipeline.  We apply two corrections to the spectra.  To correct for dust extinction in the Milky Way, we deredden the spectra using a Seaton extinction law \citep{seaton79} with $E(B-V)=0.01$ taken from the maps of \citet{schlegel98}.  We then shift the spectrum to the rest frame using the SDSS pipeline redshift of $z=0.246$. 

In addition to the SDSS photometry in the dim state, there are infrared data available from the {\it Wide-field Infrared Survey Explorer} \citep[WISE,][]{wright10} ALLWISE multi-epoch photometry catalog.  J1011+5442 was observed to have magnitudes of $W1=13.99\pm0.03$, $W2=12.86\pm0.06$, $W3=9.7\pm0.1$, and $W4=7.2\pm0.2$ on April 28, 2010 and $W1=14.19\pm0.09$ and $W2=13.08\pm0.08$ on November 5, 2010, where we list standard deviation of observations within days of each other as the magnitude uncertainty.  These data do not appear to offer any additional information on the the variability properties of this object.  J1011+5442 has not been observed by the {\it Chandra X-ray Observatory} or {\it XMM-Newton} and it was not detected in the ROSAT all-sky survey \citep{voges99}.

\begin{figure}
\rightline{\includegraphics[width=9.5 truecm]{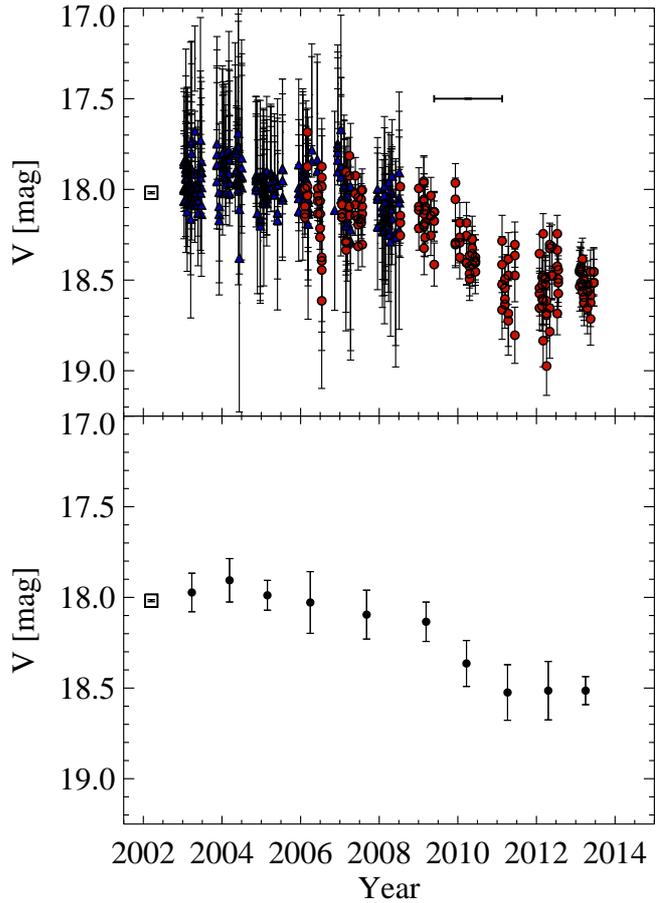}}
\caption{Top: The light curve of J1011+5442, including photometry from the CSS (solid red circles), LINEAR (solid blue triangles), and the SDSS (open black square).  The $V$ magnitude inferred from the TDSS spectrum, observed on MJD~57073, is $19.350\pm0.002$ and is not plotted here .  This value is almost certainly a lower limit on the true brightness of the AGN and host, as discussed in Section~\ref{sec:data}.  The horizontal bar represents the likely dimming timescale, $\sim$500~days in the rest frame.  Bottom: The median light curve, where each point represents the median of the observations in one year bins and the error bars represent the rms scatter in each bin.}
\label{fig:lc}
\end{figure}

J1011+5442 was observed regularly between December 2002 and June 2008 as part of the MIT Lincoln Laboratory Lincoln Near-Earth Asteroid Research (LINEAR) survey \citep{stokes00}.  This survey is performed using a single, broad filter, so we adopt the transformation of \citet{sesar11} to recalibrate the LINEAR photometry:

\begin{eqnarray}
\nonumber g &=& g_{LINEAR}-0.0574-0.004\,(g-i) \\
&+&0.056\,(g-i)^2-0.052\,(g-i)^3+0.0262\,(g-i)^4.
\end{eqnarray}

\noindent To apply the above equation, we use the SDSS $g-i$ color and convert the light curve to the $V$ band using the SDSS $g$ magnitude and the \citet{jester05} transformations.  The LINEAR light curve is shown in Figure~\ref{fig:lc} and discussed in more detail below.

Overlapping the LINEAR coverage, J1011+5442 was also observed between January 2006 and May 2013 as part of the Catalina Sky Survey \citep[CSS,][]{drake09}.  Although the observations for this survey are made in white (i.e. unfiltered) light, the light curves are reported in $V$-band magnitudes.  We further refine the reported magnitudes using the listed transformation as a function of the $V-R$ color:
\begin{equation}
V = V_{CSS} +0.91\,(V-R)^{2}+0.04.
\end{equation}
This transformation was selected from the three available because at the redshift of this quasar neither filter includes the strongest emission line, \Ha, while both filters include the second strongest emission line, \Hb, in the spectral range where they overlap. Therefore, the $V-R$ color suffers the least from emission-line contamination, compared to other colors we could have used. 

The total light curve, including the LINEAR and CSS photometry, for J1011+5442 is shown in Figure~\ref{fig:lc}. This figure also includes the SDSS magnitude, which matches the the LINEAR magnitudes and early-time measurements in the CSS light curve fairly well. The late-time TDSS fiber magnitude is $V=19.350\pm0.002$, but this corresponds to a  2\farcs0 aperture, which excludes some of the light from the extended emission from the host galaxy. Therefore, we regard this value as a lower limit on the brightness of the AGN and host and do not include it in Figure~\ref{fig:lc}.  It is not unusual to see variability in a quasar light curve; variability on the order of tenths of a magnitude is typical for quasars on timescales of a year \citep[e.g.,][]{macleod12}.  However, the systematic decline of the light curve by half of a magnitude, as displayed in Figure~\ref{fig:lc}, over a timescale shorter than several years is not typical behavior.  

From the light curve, we can estimate the timescale for the quasar flux to decay.  During the first and last years of observation, the average brightness of the quasar appears constant.  We estimate the transition timescale by visually determining where the light curve departs from the bright state and arrives at the dim state.  The poor sampling and noise in the light curve make it difficult to determine exactly when the dimming occurs, but our best estimate is $\sim$500 days in the quasar rest frame.  The relevant portion of the light curve is marked in the figure by a horizontal bar.

\subsection{Spectral decomposition}
\label{sec:specfit}
We decompose the spectra interactively using the IRAF task \textsc{specfit} \citep{kriss94}, which fits multi-component models to spectra by minimizing the $\chi^2$ statistic.  Components for the fit include a power law of the form $f_{\lambda}=A \lambda^{-\alpha}$ to characterize the quasar continuum, Gaussians to characterize the emission lines, optical \citep{bg92} and UV \citep{vestergaard01} \FeII\ templates that have been combined into a single template that covers our full wavelength range, and Charlot \& Bruzual starlight templates \citep[see][]{cales13}.  The spectral decomposition is performed over the entire available wavelength range (3100--8000~\AA\ in the dim state and 3100--7500~\AA\ in the bright state) and all components are fitted simultaneously.

\begin{figure*}
\begin{center}
\includegraphics[width=17.5 truecm]{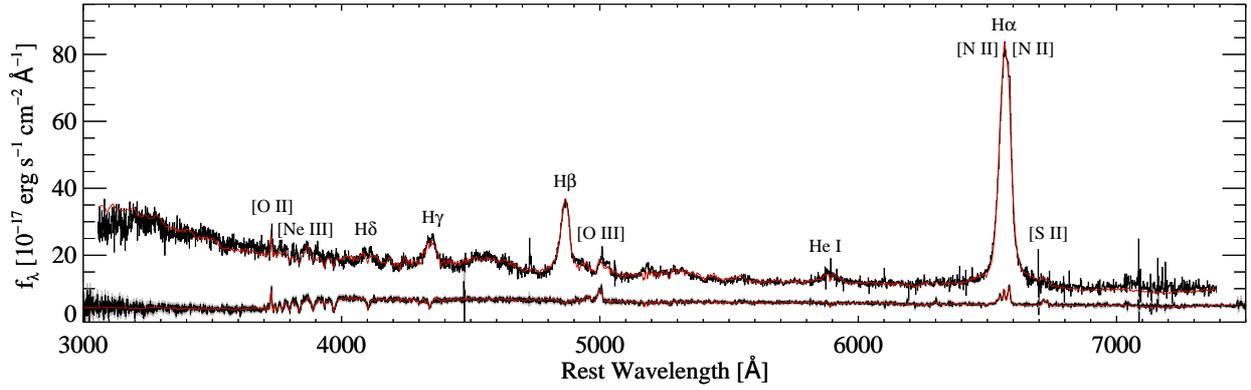}
\end{center}
\caption{The spectral fits for the bright-state (MJD~52652) and dim-state (MJD~57073) spectra.  The data are shown in black and the best-fitting total model is shown in red.  See Section~\ref{sec:data} for a description of the spectral decomposition and Figure~\ref{fig:sfitzoom} to see the flux uncertainties.}
\label{fig:fullfit}
\end{figure*}

\begin{figure*}
\begin{minipage}[!b]{8cm}
\centering
\includegraphics[width=8.9cm]{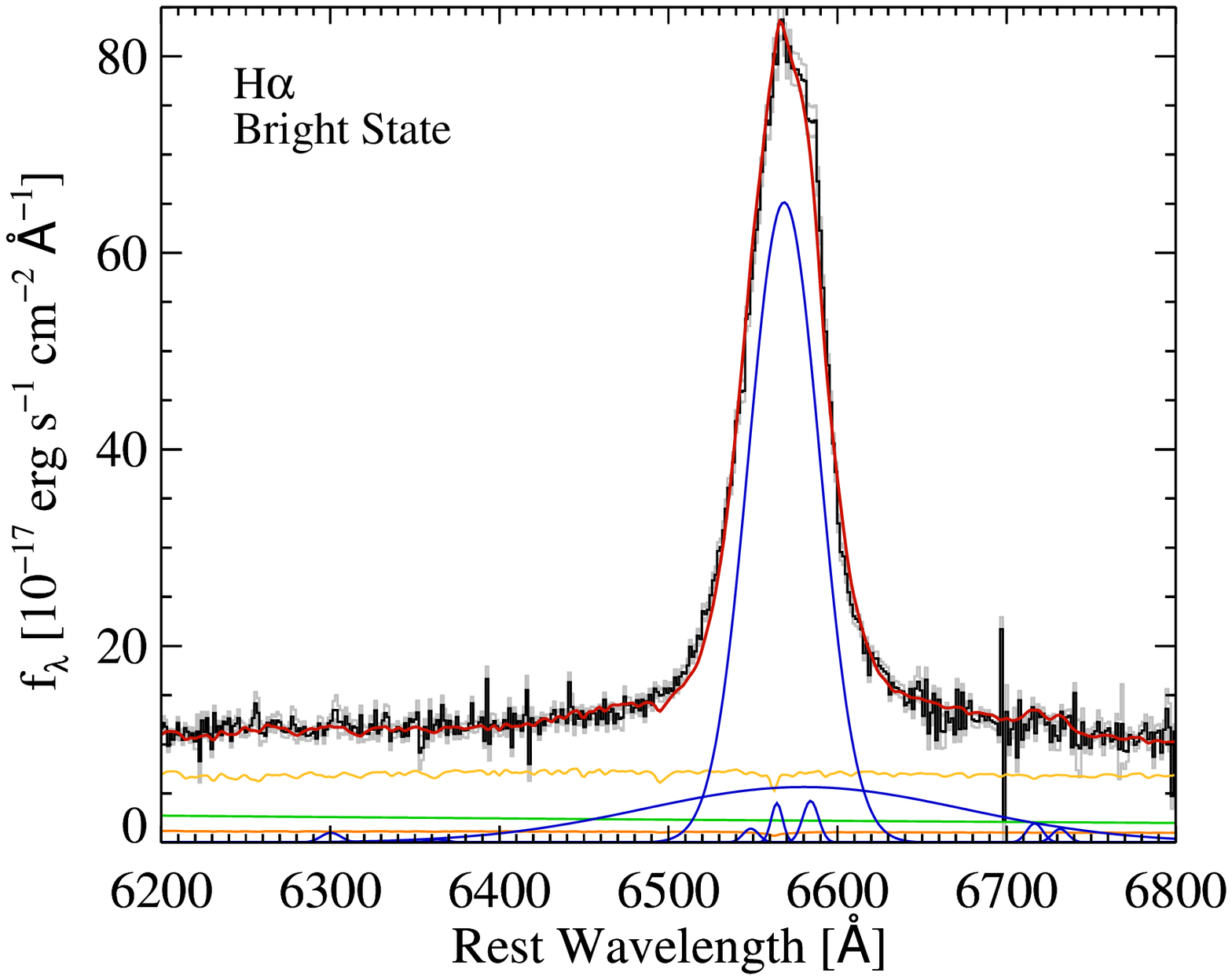}
\end{minipage}
\hspace{0.1cm}
\begin{minipage}[!b]{8cm}
\centering
\includegraphics[width=8.9cm]{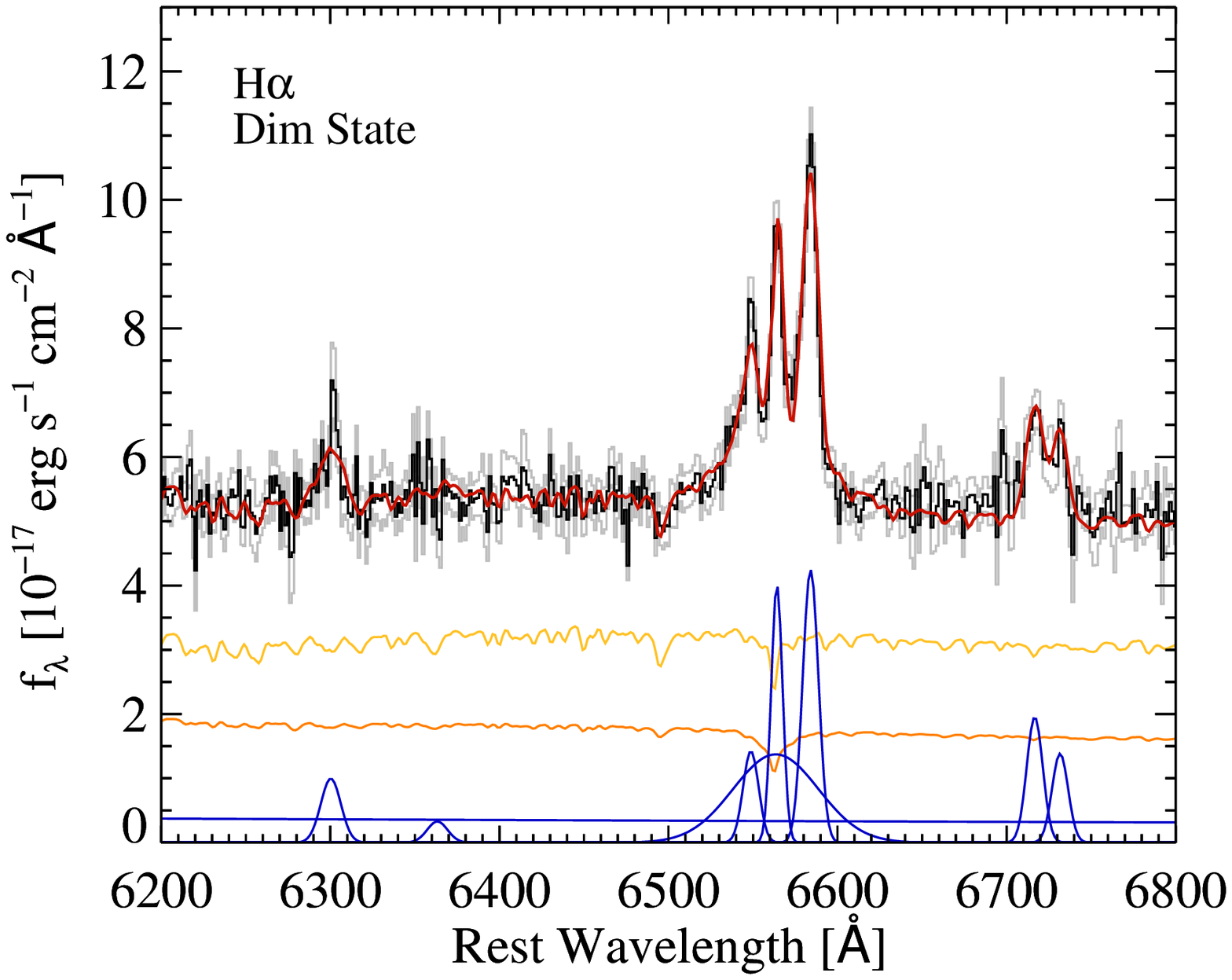}
\end{minipage}
\hspace{0.1cm}
\begin{minipage}[!b]{8cm}
\centering
\includegraphics[width=8.9cm]{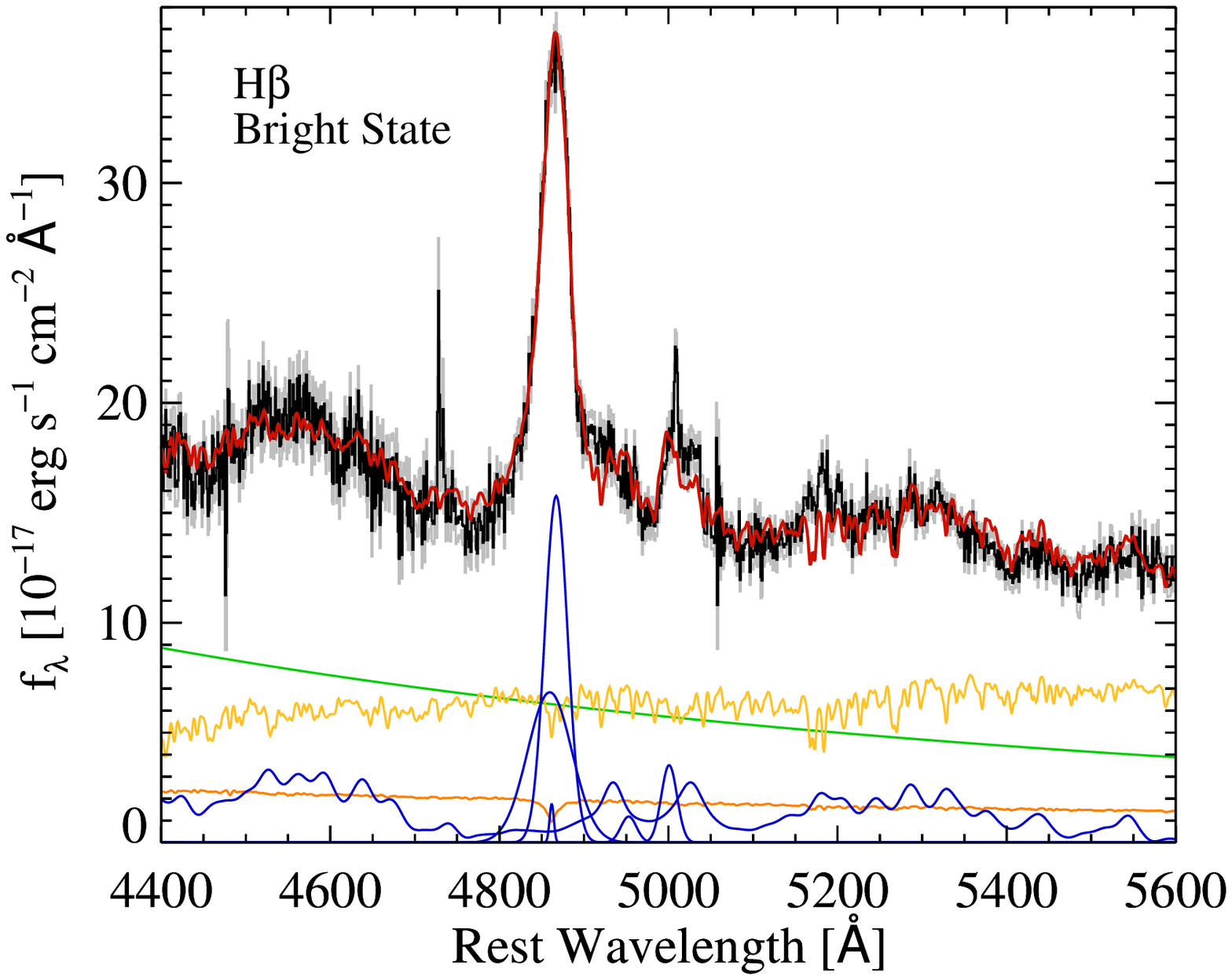}
\end{minipage}     
\hspace{0.1cm} 
\begin{minipage}[!b]{8cm}
\centering
\includegraphics[width=8.9cm]{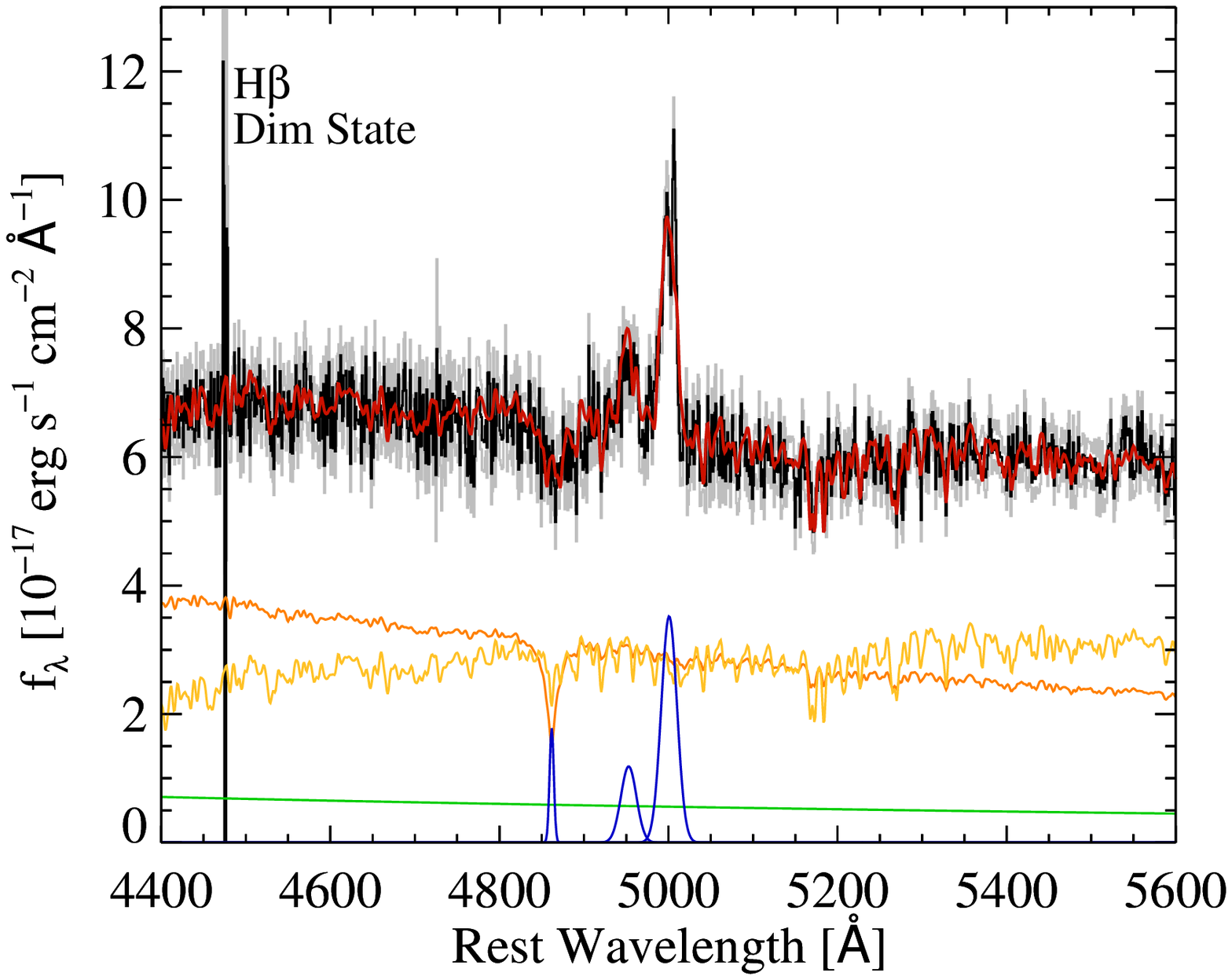}
\end{minipage}           
\begin{minipage}[!b]{8cm}
\centering
\includegraphics[width=8.9cm]{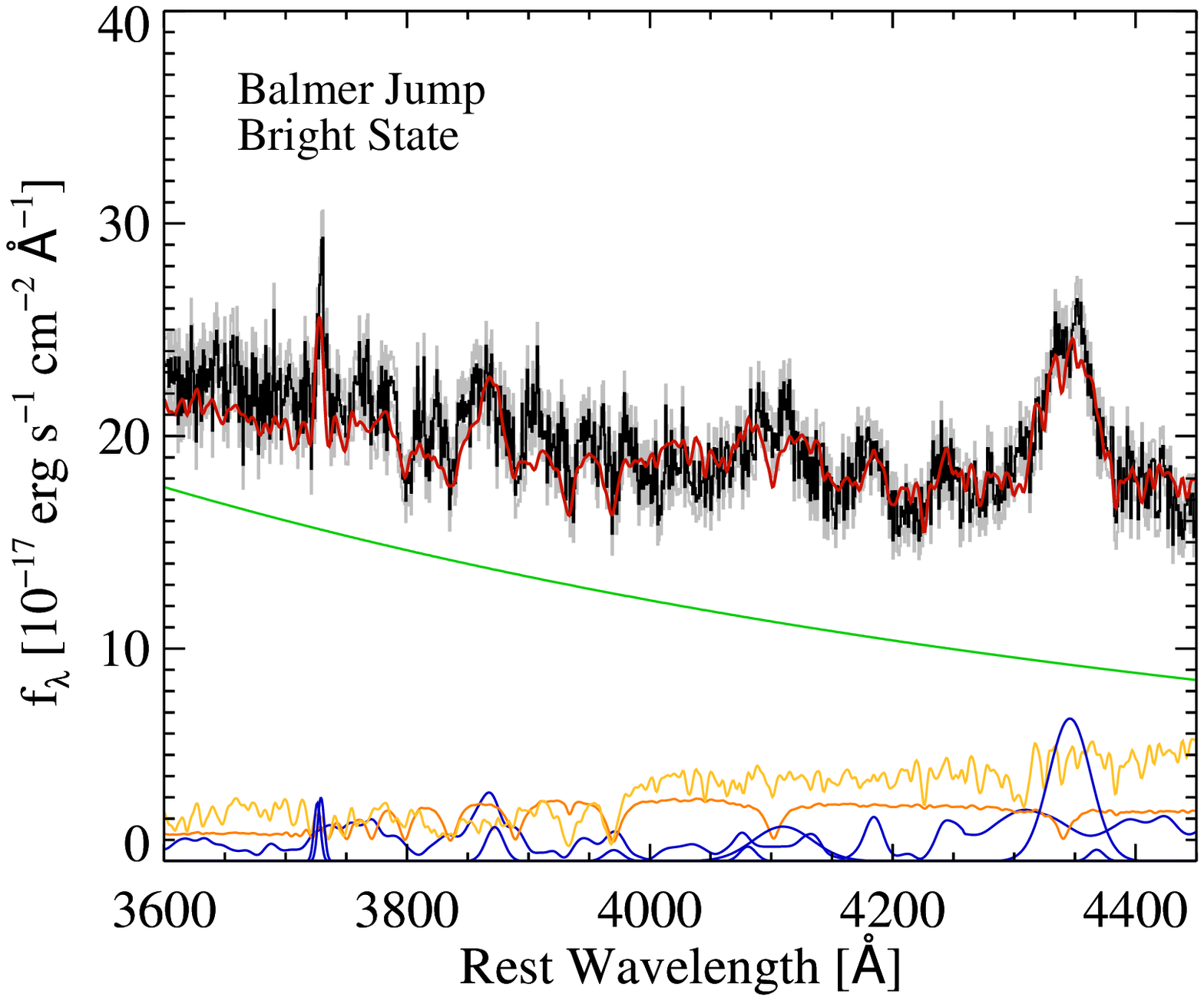}
\end{minipage}
\hspace{0.1cm}
\begin{minipage}[!b]{8cm}
\centering
\includegraphics[width=8.9cm]{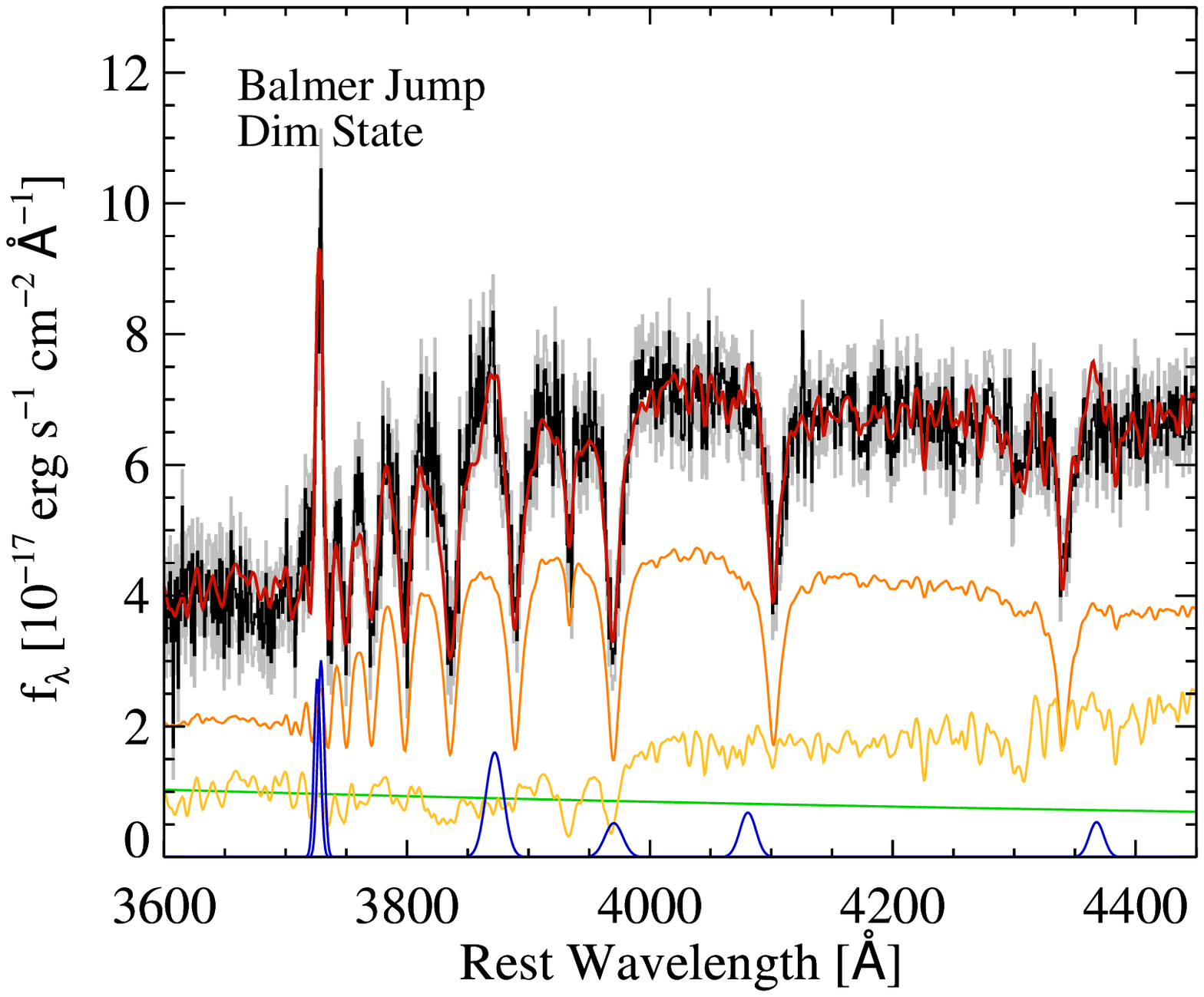}
\end{minipage}      
\caption{The spectral decomposition in three wavelength windows including \Ha\ (top), \Hb\ (middle), and strong stellar absorption lines (bottom).  The data (black), uncertainties (gray), best-fitting model (red), old and young stellar templates (yellow, orange), power-law continuum (green), and emission lines (blue, see Table~\ref{tab:measurements} for a list) are displayed for the bright state (MJD~52652) on the left and the dim state (MJD~57073) on the right.} \label{fig:sfitzoom} 
\end{figure*}

First, we decompose the dim-state spectrum because this spectrum is simpler than the bright-state spectrum. The starlight and narrow-line properties inferred from this spectrum aid us in the decomposition of the bright-state spectrum, as we explain below. We impose the following constraints on Gaussians representing narrow lines: within a doublet, the two emission lines are constrained to have identical velocity widths, a flux ratio of 3 when appropriate, and a constant wavelength ratio.  We further group the lines by ionization and critical density and constrain the velocity widths to be the same across a group; \Ha\ and \Hb\ form the first group, \OIIdblt, \NIIdblt, and \SIIdblt\ the second, and \NeIIIdblt\ and \OIIIdblt\ the last.  Finally, we also tie the wavelength of the narrow \Hb\ to that of the narrow \Ha, because the \Ha\ line is clear in the spectrum and the \Hb\ line from the quasar is weak enough that it is severely contaminated by the stellar lines.  In order to achieve a satisfactory fit, this spectrum requires two stellar populations: one where the age is fixed at 10~Gyr and the normalization is a free parameter, plus a second where both the age and normalization are allowed to vary.  This approach is similar to \citet{canalizo13}, who find significant contributions of intermediate plus old stellar populations in quasar host galaxies.  The broad lines (including the \FeII) in this spectrum are very weak; only a symmetric broad \Ha\ profile is visible and it is well characterized by a single Gaussian.  The resulting fits are shown in Figures~\ref{fig:fullfit} and \ref{fig:sfitzoom}.

In the bright state, we fit the spectrum with all the components used in the dim state, plus the \FeII\ template and a suite of Gaussians to characterize the broad lines. The broad \Ha\ and \Hb\ lines are represented by two Gaussians each in order to describe their asymmetries, while all other broad lines are represented by single Gaussians.  Since the size of the NLR \citep[$R_{NLR}\sim2.5$~kpc based on the dim-state \OIII\ luminosity and the correlation of][]{hainline13} should fall within the TDSS fiber, we fix all narrow-line properties to the best-fitting values from the dim-state decomposition.  For the starlight templates, we fix the ages to the best-fitting values from the dim state but allow the normalizations to vary.  Although we do not expect intrinsic variability from the host galaxy light, the size of the fibers used to obtain the observations were 3\farcs0 and 2\farcs0 in the bright and dim states, respectively, which creates the potential for differences between the spectra.  Visual inspection of the SDSS image of this object reveals that the host galaxy extends beyond the diameter of the fibers, suggesting that the larger fiber may include more host-galaxy light.  Thus, we do not require that the normalization of the stellar templates be identical in the two states.  

The uncertainties in all parameters resulting from the spectral decomposition are determined via a full Monte-Carlo simulation of the fitting procedure.  We resample each spectrum by selecting for each pixel a new value for the flux density, drawn from a Gaussian distribution that has a mean equal to the measured flux density in that pixel, and a standard deviation equal to the measured uncertainty in the flux density.  The best-fitting parameters for the observed spectrum are used as an initial guess, and the synthetic spectrum is refitted.  This process is repeated 10$^{3}$ times each for the bright- and dim-state spectra to generate distributions of all spectral fit parameters.  The standard deviations of these distributions are taken to be the $1\sigma$ uncertainties in the measured properties.

The complete model fits to the bright- and dim-state spectra are shown in Figure~\ref{fig:fullfit}. In Figure~\ref{fig:sfitzoom} we show an expanded view of three spectral windows of particular interest. In the latter figure we also illustrate the individual model components.  The bright and dim state emission-line measurements are listed in Table~\ref{tab:measurements}; the continuum properties we describe here.  In the bright state, the best-fitting power law has a normalization of $(1410\pm60)\times10^{-17}$~erg~s$^{-1}$~cm$^{-2}$~\AA$^{-1}$ at 1000~\AA\ and an exponent of $3.42\pm0.06$.  In the dim state, the best-fitting normalization and exponent are $(11\pm2)\times10^{-17}$~erg~s$^{-1}$~cm$^{-2}$~\AA$^{-1}$ and $1.9\pm0.3$, respectively.  The full width at half maximum (FWHM) of the lines in the \FeII\ complex, which is only present in the bright state, is $1500\pm200$~km~s$^{-1}$.  The age of the young stellar template is $225\pm4$~Myrs in both states.  The flux ratio of the young to old stellar templates at 5500~\AA\ is $0.21\pm0.01$ in the bright state and $0.76\pm0.01$ in the dim state.  We attribute the difference in this ratio to the difference in seeing conditions and fiber diameters and interpret it to mean that the young stellar population is more centrally concentrated than the old stellar population.  In the dim state, the quasar continuum is detected above the $1\sigma$ level of the noise in the data at wavelengths shorter than approximately 6800~\AA\ and the broad \Hb\ line is not detected at all.  We can place a limit on the dim-state broad \Hb\ line by asking what broad \Hb\ fluxes could be hidden in the noise in the optical part of the spectrum; using this criterion the broad \Hb\ integrated flux must be less than $13.7\times10^{-17}$~erg~s$^{-1}$~cm$^{-2}$ at the $1\sigma$ level. 

Motivated by the apparently weak power-law continuum in the dim state, we also test our model assumptions for the dim-state fit by asking whether we can obtain a similarly good decomposition without a power-law component.  We perform a second fit to the dim-state spectrum, identical to our adopted dim-state decomposition described above but with the power-law component removed.  Visually, the result describes the data well.  We compare the Bayesian Information Criterion \citep[BIC][]{schwartz78}, similar to the chi-squared statistic but appropriate for a comparison of models with a different number of free parameters, for this new fit and our adopted dim-state fit and find that statistically they describe the data equally well.  The emission line components of the fit are essentially unchanged compared to our adopted fit, so the result of removing the power-law continuum is that the normalizations of the stellar continuum components increase and the age of the young stellar population decreases to $\sim100$~Myrs.  Thus, the best-fitting power-law continuum in the dim state and the properties derived from it (e.g., luminosities) should be considered upper limits.

\begin{table*}
\begin{minipage}[2cm]{12cm}
\caption{Properties of strong emission lines \label{tab:measurements}}
\renewcommand{\thefootnote}{\alph{footnote}}
\begin{tabular}{rccccccccc}
     & & \multicolumn{3}{c}{Bright State} & & \multicolumn{3}{c}{Dim State}      \\
 & & $\lambda$ & FWHM & Integrated & & $\lambda$ & FWHM & Integrated \\
Line\footnotemark[1]     & & [\AA]                     & [km~s$^{-1}$] & Flux\footnotemark[2] & & [\AA]                     & [km~s$^{-1}$] & Flux\footnotemark[2]\\
\hline
\OII\ n\phantom{\footnotemark[2]1} & & \phantom{\footnotemark[5]}\nodata \footnotemark[5] & \nodata & \nodata & & 3729.0$\pm$0.2\phantom{\footnotemark[6]} & 499$\pm$7\footnotemark[6] & 20$\pm$1 \\
\NeIII\ n\phantom{\footnotemark[2]1} & & \nodata & \nodata & \nodata & & 3970.0$\pm$1.0\phantom{\footnotemark[6]} & 1310$\pm$60\footnotemark[6] & 10$\pm$1 \\
{H}\,$\delta$  b\phantom{\footnotemark[2]1} & & 4110.0$\pm$1.0 & 4170$\pm$60 & \phantom{1}99$\pm$6 & & \nodata & \nodata & \nodata \\
{H}\,$\gamma$ b\phantom{\footnotemark[2]1} & & 4346.0$\pm$0.7 & 2800$\pm$40 & 290$\pm$3 & & \nodata & \nodata & \nodata \\
\Hb\ b1\footnotemark[3] & & 4867.0$\pm$0.4 & 1960$\pm$60 & \phantom{1}530$\pm$40 & & \nodata & \nodata & \nodata \\
\phantom{\Hb\ }b2\footnotemark[3] & & 4860.0$\pm$0.9 & \phantom{1}3800$\pm$200 & \phantom{1}450$\pm$40 & & \nodata & \nodata & \nodata \\
\phantom{\Hb\ }b\footnotemark[4]\phantom{1} & & 4864.0$\pm$0.3 & \phantom{1}2350$\pm$50\phantom{1} & \phantom{1}980$\pm$10 & & \nodata & \nodata & <13.7 \\
\phantom{\Hb\ }n\phantom{\footnotemark[2]1} & & \nodata & \nodata & \nodata & & 4861.9$\pm$0.1\footnotemark[6] & 361$\pm$7\footnotemark[6] & 11$\pm$1 \\
\OIII\ n\phantom{\footnotemark[2]1} & & \nodata & \nodata & \nodata & & 5001.0$\pm$0.4\phantom{\footnotemark[6]} & 1310$\pm$60\footnotemark[6] & 82$\pm$3 \\
He\,{\sc i} b\phantom{\footnotemark[2]1} & & 5895.0$\pm$1.0 & \phantom{1}3200$\pm$300 & \phantom{1}200$\pm$10 & & \nodata & \nodata & \nodata \\
\OI\ n\phantom{\footnotemark[2]1} & & \nodata & \nodata & \nodata & & 6363.0$\pm$0.5\phantom{\footnotemark[6]} & \phantom{1}670$\pm$30 & \phantom{1}5$\pm$1 \\
\NII\ n\phantom{\footnotemark[2]1} & & \nodata & \nodata & \nodata & & 6584.0$\pm$0.1\phantom{\footnotemark[6]} & 499$\pm$7\footnotemark[6] & 50$\pm$2 \\
\Ha\ b1\footnotemark[3] & & 6568.0$\pm$0.1 & \phantom{1}2260$\pm$20\phantom{1} & 3430$\pm$40 & & 6563.5$\pm$0.1\phantom{\footnotemark[6]} & 2720$\pm$70 & 87$\pm$3 \\
\phantom{\Ha\ }b2\footnotemark[3] & & 6580.0$\pm$1.0 & 10200$\pm$400 & 1340$\pm$30 & & \nodata & \nodata & \nodata \\
\phantom{\Ha\ }b\footnotemark[4]\phantom{1} & & 6571.0$\pm$0.0 & \phantom{1}2420$\pm$30\phantom{1} & 4770$\pm$30 & & 6563.5$\pm$0.1\phantom{\footnotemark[6]} & 2720$\pm$70 & 87$\pm$3 \\
\phantom{\Ha\ }n\phantom{\footnotemark[2]}\phantom{1} & & \nodata & \nodata & \nodata & & 6564.0$\pm$0.1\footnotemark[6] & 361$\pm$7\footnotemark[6] & 34$\pm$1 \\
\SII\ n1\phantom{\footnotemark[2]} & & \nodata & \nodata & \nodata & & 6717.0$\pm$0.6\footnotemark[6] & 499$\pm$7\footnotemark[6] & 23$\pm$1 \\
\phantom{\SII\ }n2\phantom{\footnotemark[2]} & & \nodata & \nodata & \nodata & & 6732.0$\pm$0.6\footnotemark[6] & 499$\pm$7\footnotemark[6] & 17$\pm$1 \\
\hline
\end{tabular}
\footnotetext[1]{b = broad component (number identifies Gaussian components summed to produce total line profile), n = narrow component.  Wavelengths for lines that are in a doublet are for the long-wavelength line.}
\footnotetext[2]{Integrated fluxes are in units of 10$^{-17}$ erg s$^{-1}$ cm$^{-2}$.}
\footnotetext[3]{One of two Gaussian components.}
\footnotetext[4]{Total profile, the sum of two Gaussian components.}
\footnotetext[5]{The bright-state narrow lines parameters are fixed to the dim-state values.}
\footnotetext[6]{This value is constrained during the spectral decomposition.  See Section~\ref{sec:specfit} for a description of the constraints.}
\end{minipage}
\end{table*}

\subsection{Inferred quasar properties}
Using the decomposed bright-state spectrum, we measure fundamental properties of the central black hole.   In the bright state, we apply the 5100~\AA\ bolometric correction with a nonzero intercept from table~1 of \citet{runnoe12c} to the monochromatic optical luminosity $\lambda\,L_{\lambda}(5100\,\textrm{\AA})=(6.3\pm0.1)\times10^{43}$~erg~s$^{-1}$ to calculate a bolometric luminosity of $L_{bol}=(6.8\pm0.1)\times10^{44}$~erg~s$^{-1}$. The broad \Hb\ FWHM of $2350\pm50$~km~s$^{-1}$, combined with the single-epoch mass scaling relationship from \citep{vestergaard06}, yields $M_{BH}=(3.6\pm0.2)\times 10^{7}\,M_{\odot}$.  These estimates produce an Eddington ratio of $L/L_{Edd} = 0.15$.  We can also calculate the black hole mass from \Ha\ by adopting the scaling relationship from \citep{greene10a}.  This calibration differs slightly from that of \citet{vestergaard06}, as discussed by \citet{cales13}, but the difference is small compared to the intrinsic scatter.  This approach yields $M_{BH}=(4.7\pm0.1)\times 10^{7}\,M_{\odot}$ and $L/L_{Edd} = 0.12$.  The uncertainties quoted on these mass measurements do not include the $0.4-0.5$~dex scatter associated with the single-epoch mass scaling relationships.  This additional source of error translates to a factor of $2.5-3$ uncertainty in mass and is more representative of the uncertainty than the formal errors.  Thus, the \Ha\ and \Hb\ masses are consistent with each other given the large uncertainties associated with the mass scaling relationships.  Using the same relationships for the dim state, $\lambda\,L_{\lambda}(5100\,\textrm{\AA})=(6.3\pm0.1)\times10^{42}$~erg~s$^{-1}$, $L_{bol}=(8.4\pm0.1)\times10^{43}$~erg~s$^{-1}$, $M_{BH}=(1.8\pm0.3)\times 10^{7}\,M_{\odot}$, and $L/L_{Edd} = 0.04$ (the mass measurement is for \Ha\ as \Hb\ is not detected).

In principle, because the black hole mass is constant, the continuum luminosity and broad-line widths should change in concert such that the single-epoch mass measurements are consistent between the bright and dim states.  \citet{lamassa15} found this to be the case for J0159+0033.  Our estimates for the \Ha\ masses are preserved within a factor of three, although not within the formal $1\sigma$ uncertainties.

\begin{figure}
\begin{center}
\includegraphics[width=8.9 truecm]{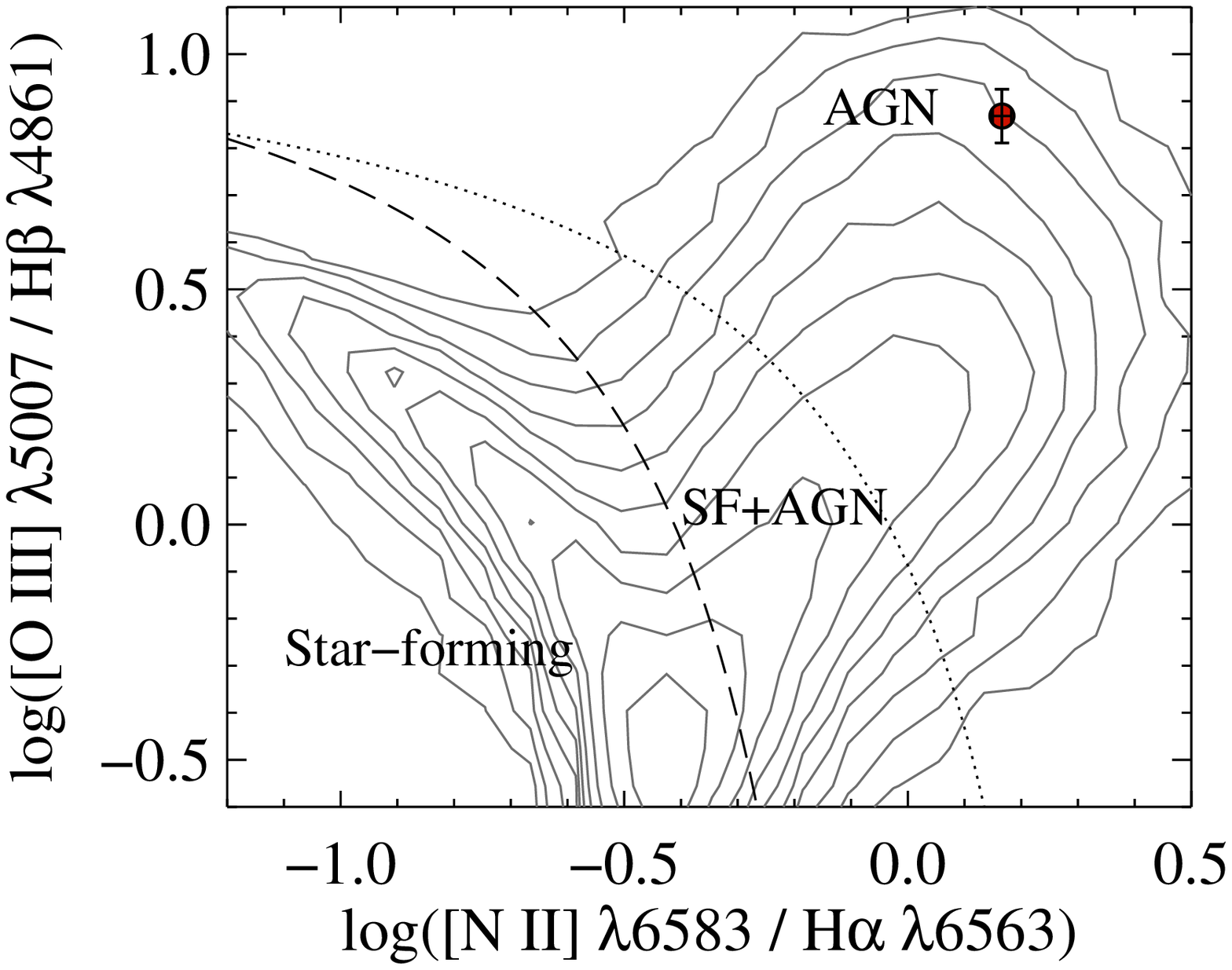}
\end{center}
\caption{Diagnostic emission line ratios for J1011+5442.  The narrow lines in the bright-state spectral decomposition were fixed to the dim-state values, so only one position is obtained on the BPT diagram.  The BPT identification schemes from \citet{kewley01} and \citet{kauffmann03} are shown as dotted and dashed lines, respectively.  For comparison, gray contours show the properties of all SDSS galaxies in \citet{tremonti04}.  The emission-line ratios in J1011+5442 are consistent with a pure AGN ionizing continuum.}
\label{fig:bpt}
\end{figure}

To constrain the excitation mechanism of the narrow-line gas, we plot the observed \OIII/\Hb\ and \NII/\Ha\ narrow-line ratios in a diagnostic diagram \citep*[also known as a ``BPT'' diagram after][]{baldwin81}. The location of the measured ratios in this diagram can reveal the relative contributions to the ionizing continuum from the AGN and star formation.  The line ratios from the dim-state spectrum, where the narrow-line decompositions as performed, are shown in Figure~\ref{fig:bpt}.  We include the classification schemes of \citet{kewley01} and \citet{kauffmann03} to separate line ratios characteristic of AGN, star-forming, and composite galaxies and show as contours the distribution of SDSS galaxies from \citet{tremonti04} for comparison.  The conclusion for J1011+5442 is clear: a hard AGN continuum is required to produce the observed narrow lines.  This does not change if we adopt the alternative line ratios of \OIw/\Ha\ or \SIIdblt/\Ha\ in place of \NII/\Ha.  

\section{Analysis}
\label{sec:analysis}
In this section, we perform a series of tests on the spectra and LINEAR+CSS light curve in order to constrain the physical mechanism responsible for the drastic changes observed in J1011+5442. We consider three physical scenarios for the observed dimming of the quasar central engine: abrupt obscuration of the central engine, intrinsic dimming due to a drop in the accretion rate \citep[also examined as explanations for the dimming of J0159+0033 by][]{lamassa15}, and decay of a flare produced by the tidal disruption of a star by the central BH \citep[proposed in the context of J0159+0033 by][]{merloni15}.

\subsection{Obscuration of the central engine}
\label{sec:reddening}
Following \citet{lamassa15}, we examine whether the time required for a foreground object in a bound orbit around the BH to move in front of the continuum source and BLR can be short enough to match the observed transition time.  We adopt equation~4 of \citet{lamassa15}, 

\begin{equation}
\label{eqn:tcross}
t_{cross} = 0.07 \left[ \frac{r_{orb}}{1\,\textrm{lt} - \textrm{day}}\right]^{3/2} M_{8}^{-1/2} \textrm{arcsin}\left[ \frac{r_{src}}{r_{orb}}\right]\textrm{yr},
\end{equation}

\noindent which gives the time for an object at an orbital radius $r_{orb}$ to traverse a region extending to radius $r_{src}$ around a black hole of mass $M_{8}=M/10^8M_{\sun}$.  The most conservative case, which minimizes the crossing time, corresponds to the minimum plausible value of $r_{src}$ and a value of $r_{orb}$ that is determined graphically. Thus, we begin by assuming that $r_{src}$ is the radius derived from the ``Clean2+ExtCorr'' radius-luminosity relation \citep[$R_{BLR}$,][]{bentz13}, i.e., an emissivity-weighted radius that encompasses the bulk of the broad-line emission. For J1011+5442 $R_{BLR}=28$ light-days; assuming that $r_{src}=R_{BLR}$, we estimate a minimum crossing time of $t_{cross}\sim21$~years with $r_{orb} = 34$ light days. This minimizes the crossing time since we have effectively assumed that the obscuring object is orbiting within the BLR and is large enough to obscure its brightest parts.  The absolute lower limit on the crossing time is $t_{cross}\sim9$~years, calculated by subtracting the 0.13~dex scatter in the $R-L$ relationship from log$(R_{BLR})$, increasing the mass by a factor of 2.5 to account for the uncertainty in the mass scaling relationship, and recalculating the crossing time under the same assumptions about $r_{src}$ and $r_{orb}$.  However, more realistically, the obscuring object should orbit outside the BLR, whose outer radius is several times larger than $R_{BLR}$. Reverberation mapping of infrared emission from illuminated dust indicates that the BLR is contained within $\sim3\,R_{BLR}$ \citep[e.g.,][see \citealt{lamassa15} for a more detailed justification of this assumption]{barth13,suganuma06,baldwin04}, so we set $r_{orb}=r_{src}=3\,R_{BLR}$.  This leads to a crossing time of $t_{cross}\sim31$~years.  To obscure all of the BLR, or for screens located farther from the source, crossing times can easily reach hundreds of years.  In all cases, these timescales are far too long to explain the dimming seen here.  

If the power-law continuum is detected, we can carry out a complementary test by comparing the ratio of fluxes from the dim- and bright-state spectra at specific wavelengths and asking whether they are all consistent with the same amount of reddening.  The ratio of the \Ha\ flux between the dim and bright states is $0.018\pm0.001$, which requires $E(B-V)=1.744$ in order to be explained by extinction.  The flux ratio of \Ha/\Hb\ changes by $>1.3$ from the bright to dim state, implying $E(B-V)>0.248$, which is consistent with the decline in the \Ha\ flux.  The ratio of the quasar continua from the two epochs at 6563~\AA\ is $0.15\pm0.08$.  which is not consistent with the amount of dimming measured at the same wavelengths in the \Ha\ line.  The \Hb\ line and nearby continuum should also demonstrate the same amount of attenuation, but this line ratio is $<0.014$ whereas the 5100~\AA\ continuum is $0.100\pm0.003$.  Finally, at 3500~\AA\ the amount of reddening needed to attenuate \Ha\ produces a continuum flux that is about a factor of 100 too weak.  Thus, if the continuum is detected from the quasar the relative dimming of the emission lines and the continuum is not consistent with reddening.  If, on the other hand, the power-law continuum is an upper limit, this discrepancy is alleviated and the real problem for the reddening scenario is the timescale for obscuration to set in.

Additional evidence against the reddening scenario can be found in the broad \Ha\ line width; the line profile changes such that \Ha\ is broader in the dim state, suggesting that it is emitted nearer to the central engine which would presumably suffer from the most extinction.  Based on all these calculations, we disfavor reddening by a foreground screen as an explanation of the changes in this object.

\subsection{Intrinsic dimming of the continuum source}
\label{sec:calc}
An alternative explanation for the observed variability is that a rapid change in accretion rate is responsible for the dimming in the continuum and broad emission lines.  We evaluate whether this might be possible by calculating the accretion timescale for the inner part of the accretion disk where most of the ionizing radiation is generated ($R<10\,R_{g}$, where $R_g=GM_{BH}/c^2$ is the gravitational radius).  We follow equation~5 of \citet{lamassa15} for the inflow timescale (i.e., the time for a parcel of gas to move radially from a given radius in the accretion disk to the center, also known as the viscous or radial drift timescale), adopting their fiducial value for the viscosity parameter \citep{shakura73} but substituting our measured values of $M_{8}=0.36$ and $\lambda_{Edd} = 0.15$.  This gives an inflow time of $t_{infl} = 0.25$~years, or equivalently three months.  This timescale is short enough that a drop in accretion can easily explain the changes in the continuum that are observed in J1011+5442. The inflow time scale we estimate here is considerably shorter than what \citet{lamassa15} estimated for J0159+0033 because the two quasars have considerably different masses and accretion rates (in the bright state, J1011+5442 has $M_{8}=0.36$ and $\lambda_{Edd} = 0.15$ versus $M_{8}=1.7$ and $\lambda_{Edd} = 0.04$ for J1059+0033).

It is interesting to note that the continuum throughout the optical band dimmed within approximately 500 days, as well as the far-UV continuum at $\sim 1\;$Ry, which is responsible for the ionization of Hydrogen and the production of the Balmer lines. As noted by \citet{lamassa15}, these changes may be driven by a drop in the luminosity of the inner accretion disk even though the optical continuum is produced in a region of the accretion disk that is much larger than $10\;R_g$. This because the optical continuum may result from illumination from the inner disk, as suggested by the results of reverberation studies \citep[e.g.,][]{krolik91,cackett07}. Similarly, since the ``radius of the broad-line region,'' $R_{BLR}$, is only 28~light-days, the luminosity of the broad Balmer emission lines will track the decline of the Lyman continuum with a lag that is less than a month. Therefore, the transformation of J1011+5442 can be accomplished by a rapid drop in the far-UV continuum luminosity resulting from an abrupt decline in the accretion rate. The mechanism behind this change in the accretion rate, however, is unknown. Tantalizing as this scenario may be, it must remain tentative until a plausible mechanism for abrupt changes in accretion rate is identified.

\subsection{Tidal disruption events}
A decaying flare from a tidal disruption event (TDE) was proposed as a possible mechanism for the changes observed in J0159+0033 by \citet{merloni15}.  Specifically, J1059+0033 was initially observed in a lowly accreting state, but brightened dramatically before spectroscopy revealed the presence of broad lines, which later disappeared.  We use the LINEAR+CSS light curve to evaluate if the TDE scenario can explain J1011+5442.  In this case, there is no dramatic brightening of the light curve, so the entire event is not covered.  Rather, the hypothetical TDE is already underway at the start of our observations so we resolve only part of the peak of the light curve followed by its decline.

For a more quantitative investigation, our first step is to derive the light curve of the variable component by subtracting the flux of the host galaxy.  Although the TDSS spectrum provides a good measure of the host emission, the 2\farcs0 aperture likely misses some of the flux.  Therefore, instead of using the TDSS spectrum, we assume that the faintest point in the CSS light curve provides a limit on the host flux, which we subtract from all points in the light curve.

\begin{figure}
\begin{center}
\includegraphics[width=8.9 truecm]{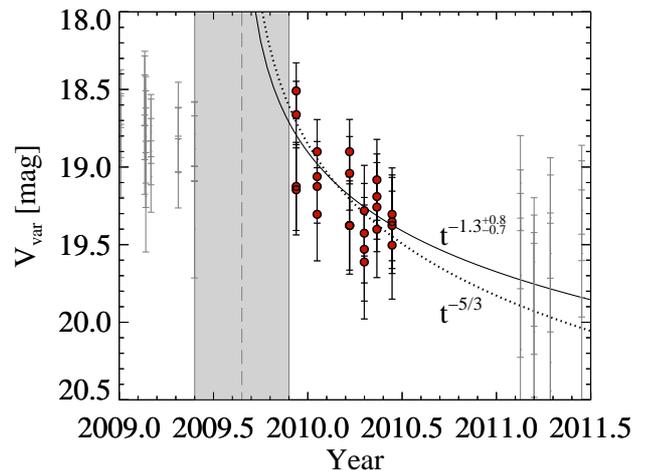}
\end{center}
\caption{The Catalina Sky Survey light curve of the variable emission with the host subtracted.  Red points indicate data that were included in fitting the TDE model, while other points are shown with gray error bars only.  The gray region indicates the observational constraints on $t_{0}$ and the dashed gray line is the adopted value.  The solid black line represents the best-fitting TDE model, where $\alpha = -1.3^{+0.8}_{-0.7}$, and the dotted line shows the same model with the slope fixed at $\alpha=-5/3$. \label{fig:qlc}}
\label{fig:css_lc}
\end{figure}

Analytic calculations \citep{phinney89} and numerical modeling \citep[e.g.,][]{evans89,ayal00} of TDEs predict that the light curve of a TDE will decline as $t^{\alpha}$, where $\alpha=-5/3$.  In principle, this $t^{-5/3}$ decay law that is traditionally associated with TDEs actually describes the rate at which material will return to the vicinity of the black hole, which is likely not equal to the accretion rate. In fact, \citet{lodato10} predict different power-law indices for light curves in different electromagnetic bands. Nonetheless, the majority of the tidal disruption events claimed to date do exhibit this decay law in their light curves \citep[e.g.,][]{gezari09}.   To test whether the declining light curve of J1011+5442 is consistent with this expectation, we write the evolution of its $V$-band magnitude as
\begin{equation}
\label{eqn:tde}
V(t) = V_{0} + \alpha\,\textrm{log}\left(\frac{t-t_{0}}{1\,\textrm{yr}}\right),
\end{equation}
where $t_{0}$ indicates the beginning of the decline in the light curve, observationally constrained to be between 2009.4 and 2009.9.  To fit the light curve, we adopt $t_{0} = 2009.65$ and use the range in $t_{0}$ to determine the uncertainty on $\alpha$.  We find $\alpha = -1.3^{+0.8}_{-0.7}$, which is formally consistent with $-5/3$; both curves are presented with the host-subtracted light curve in Figure~\ref{fig:qlc}. We note that, for some specific values of $t_0$ within 2009.4--2009.9, it is possible to obtain slopes that are inconsistent with $-5/3$.

We can also calculate the amount of matter accreted by the black hole during the bright state and compare it to reasonable values for stars that might be tidally disrupted.  That is to say, the amount of matter accreted during the high state cannot exceed the mass of the disrupted star.  Assuming an efficiency of $\eta=0.1$, and adopting the measured bolometric luminosity from the high state, the black hole accreted 0.7~\Msun\ in the $\sim$6 years that it was in the bright state.

Although the above tests appear to support the TDE interpretation, the overall shape of the light curve does not. The most recent simulations for feeding black holes with tidal disruption of stars show that the light curve of a TDE has a very narrow peak \citet{guillochon13}.  From their figure~5, a timescale of a few months is a generous upper limit on the bright peak in the light curve of a TDE.  Similarly, the observed light curves of strong TDE candidates \citep[e.g.,][]{gezari09} remain in the bright state only for a very short time, timescales on the order of months.  In contrast, J1011+5442 is in the bright state for at at least $\sim$6 years in the rest frame.

The bright-state emission lines are difficult to explain in the TDE scenario.  The observed broad emission lines suggest a BLR that is too extended and massive to have been deposited by the recent tidal disruption of a single star.  The narrow lines, which do not change, also present challenges, requiring a hard ionizing continuum and gas at even larger radii.  \citet{merloni15} discuss these points and relevant calculations in detail, but the issue remains that, while a TDE can illuminate an existing BLR, it cannot distribute the necessary gas around the black hole or produce narrow lines coincident with the peak of the light curve.  Based on all these lines of evidence, we therefore reject the TDE scenario to describe J1011+5442.  

\section{Summary and discussion}
\label{sec:discussion}
In this work, we report the discovery and analysis of the $z=0.246$ changing-look quasar, SDSS J101152.98+544206.4.  In the bright state, observed in early 2003 as part of the SDSS, J1011+5442 is listed in the SDSS quasar catalogs up to DR7 and the optical spectrum of J1011+5442 exhibits the broad emission lines and a blue continuum of a typical broad-lined quasar.  From this spectrum, we determine $L_{bol}=(6.8\pm0.1)\times10^{44}$~erg~s$^{-1}$, $M_{BH}=(3.6\pm0.2)\times 10^{7}\,M_{\odot}$, and $\lambda_{Edd}=0.15$.  The TDSS targeted the object for observations in early 2015 and found it in a dim state: the quasar continuum was undetected at the $1\sigma$ level at wavelengths longer than $\sim6800$~\AA, and of the broad emission lines, only \Ha\ was still present, but at a much weaker level than was observed during the bright state in 2003.  The 2006$-$2013 CSS light curve reveals that the transition likely occurred over $\sim500$~days in the rest frame around 2010.    

We test several physical scenarios to explain the dramatic changes in J1011+5442's spectrum.  In some cases (typically in Seyfert 1.8--1.9 galaxies), variable obscuration of the central engine can explain changing-look AGN, but this mechanism is disfavored in this case.  The timescale for an intervening cloud of dust to obscure the broad-line region is too long, on the order of 30 years, compared to the measured timescale of $\sim500$~days.  An additional test of the obscuration scenario would be to ask whether the observed change in the quasar continuum is consistent with the amount of reddening needed to produce the change in \Ha.  We attempt this test, but the results are inconclusive because the quasar continuum is not robustly detected in the optical.  {\it Hubble Space Telescope} observations in the ultra-violet, where the power-law continuum rises, would be required to perform this test conclusively.  The observations of J1011+5442 cannot be explained by a TDE.  Although the decay rate of the light curve is consistent with theoretical expectations for TDEs and the 0.7~\Msun\ accreted while J1011+5442 was in the bright state could have been deposited by disruption of a star, the bright state has a minimum duration that is far longer than any putative observed TDE or theoretical TDE model.  An abrupt change in the accretion rate can drive the changing look of J1011+5442.  The timescale for material to be transported through the inner accretion disk and for the disk to respond to changes in the accretion rate is on the order of months, consistent with the observed change, making this a viable process. 

In the context of recently discovered changing-look quasars \citep[][Ruan~et~al., submitted; Cales et al. in preparation]{lamassa15}, J1011+5442 has similar luminosity, redshift, and transition timescale. All recently-discovered changing-look quasars cluster in the upper right corner of the luminosity-redshift diagram of \citet[][see their Figure~1]{lamassa15}. J1011+5442 does, however, have a notably smaller black hole mass and larger Eddington ratio compared to J0159+0033, which lead to a much shorter inflow time scale in the inner accretion disk (months compared to decades for J0159+0033). As a result, a large, abrupt fluctuation in the accretion rate is a viable explanation of the variability of J1011+5442, whereas with J0159+0033 there is some potential for ambiguity.  \citet{elitzur14} describe an accretion-rate driven paradigm for the evolution between spectral types.  This is based on the sample of intermediate-type AGNs from \citet{stern12b}, where a decreasing fraction of the bolometric luminosity is converted to broad-line emission as the accretion rate, characterized by $L_{bol}/M_{BH}^{2/3}$, drops.  J1059+0033 is consistent with the expected values of  $L_{bol}/M_{BH}^{2/3}$ for this sequence but it is not possible to ascribe the spectral changes to changes in the accretion rate with complete certainty because the inflow timescale is so long, whereas J1011+5442 is a fairly unambiguous case because of the shorter inflow timescale.  Thus, the changing look-quasars appear to be more luminous analogs of {\it some of} the Seyfert 1.8--1.9 galaxies studied by \citet{goodrich89,goodrich90,goodrich95} and later \citet{trippe10}.  As Goodrich and Trippe~et~al. conclude from a variety of observational tests, the variability of the continuum and broad Balmer lines of some objects could be attributed to intrinsic variations while in others it was caused by changes in the reddening. 

Our understanding of changing-look quasars will benefit substantially from continued optical spectroscopic monitoring as well as X-ray observations. X-ray spectroscopy was important in rejecting the obscuration scenario for J0159+0033 and can serve the same purpose for the other changing-look quasars. Spectroscopic monitoring can serve at least two purposes. First, it can reveal whether the quasar central engine turns on again because X-ray emission originates near the black hole and acts as an instantaneous tracer of accretion.   As we discover and monitor larger numbers of such objects we will be able to constrain the duration of the ``off-state.''  Moreover, if the quasar returns, this would be a strong indication that the previous ``on-state'' was not caused by a TDE since the rate of such events is expected to be $\sim 10^{-5}$--$10^{-4}$~galaxy$^{-1}$~yr$^{-1}$ \citep[both on theoretical and observational grounds, e.g.,][]{merritt04,donley02,gezari08,maksym10}. Second, as \citet{lamassa15} remarked, we may observe a decline in the {\it narrow} emission lines, starting potentially as early as several years after the decline of the continuum.  The precise onset of the response will depend on the structure of the NLR in J1011+5442, but can happen more quickly than imagined given that the traditional kiloparsec-scale radius of the NLR describes the outer extent rather than the innermost radius of the gas \citep[see discussions on narrow-line variability in e.g.,][]{lamassa15,denney14}.  This would afford an additional test of the intrinsic dimming hypothesis and it would allow us to probe the structure of the narrow-line region. There are, in fact, specific predictions of how the narrow-lines would respond to a sharp drop in the ionizing continuum, based on time-dependent photoionization calculations by \citet{binette87} and \citet{eracleous95}. Among the noteworthy predictions are the rapid response of the optical [\ion{O}{3}] doublet (the line flux tracks the decline of the ionizing continuum with a lag of only a few years) and the very slow response of the Balmer lines (they remain fairly steady for decades to centuries after the ionizing continuum has faded).

The TDSS offers a promising way of discovering substantial numbers of changing-look quasars because it will revisit several thousand objects with previous spectra from the SDSS. The two changing look quasars discovered within the first year of the TDSS may be the tip of the iceberg. By design, however, the TDSS is unlikely to find transitions from the ``off'' to the ``on'' state, which are just as interesting and valuable in their own right, because the TDSS targets are known AGN.

\section{Acknowledgements}
Funding for the Sloan Digital Sky Survey IV has been provided by the Alfred P. Sloan Foundation and the Participating Institutions. SDSS-IV acknowledges support and resources from the Center for High-Performance Computing at the University of Utah. The SDSS web site is www.sdss.org.

SDSS-IV is managed by the Astrophysical Research Consortium for the Participating Institutions of the SDSS Collaboration including the Brazilian Participation Group, Carnegie Institution for Science, Carnegie Mellon University, the Chilean Participation Group, Harvard-Smithsonian Center for Astrophysics, Instituto de Astrofísica de Canarias, The Johns Hopkins University, Kavli Institute for the Physics and Mathematics of the Universe (IPMU) / University of Tokyo, Lawrence Berkeley National Laboratory, Leibniz Institut für Astrophysik Potsdam (AIP), Max-Planck-Institut für Astrophysik (MPA Garching), Max-Planck-Institut für Extraterrestrische Physik (MPE), Max-Planck-Institut für Astronomie (MPIA Heidelberg), National Astronomical Observatory of China, New Mexico State University, New York University, University of Notre Dame, Observatório Nacional do Brasil, The Ohio State University, Pennsylvania State University, Shanghai Astronomical Observatory, United Kingdom Participation Group, Universidad Nacional Autónoma de México, University of Arizona, University of Colorado Boulder, University of Portsmouth, University of Utah, University of Washington, University of Wisconsin, Vanderbilt University, and Yale University.

The CSS survey is funded by the National Aeronautics and Space Administration under Grant No. NNG05GF22G issued through the Science Mission Directorate Near-Earth Objects Observations Program.  The CRTS survey is supported by the U.S.~National Science Foundation under grants AST-0909182.

The LINEAR program is funded by the National Aeronautics and Space Administration at MIT Lincoln Laboratory under Air Force Contract FA8721- 05-C-0002. Opinions, interpretations, conclusions, and recommendations are those of the authors and are not necessarily endorsed by the United States Government.

This publication makes use of data products from the Wide-field Infrared Survey Explorer, which is a joint project of the University of California, Los Angeles, and the Jet Propulsion Laboratory/California Institute of Technology, funded by the National Aeronautics and Space Administration.

\bibliographystyle{./mnras}
\bibliography{./all.052615}

\label{lastpage}
\end{document}